\documentclass[%
 reprint,
 amsmath,amssymb,
 aps,
]{revtex4-2}

\usepackage{graphicx}% Include figure files
\usepackage{dcolumn}% Align table columns on decimal point
\usepackage{bm}% bold math
\usepackage[utf8]{inputenc}
\usepackage[T1]{fontenc}
\usepackage{mathptmx}
\usepackage{etoolbox}

\usepackage{amssymb}
\usepackage{float}
\usepackage[normalem]{ulem}

\DeclareMathOperator{\arctantwo}{arctan2}
\DeclareMathOperator{\sgn}{sgn}

\makeatletter
\def\@email#1#2{%
 \endgroup
 \patchcmd{\titleblock@produce}
  {\frontmatter@RRAPformat}
  {\frontmatter@RRAPformat{\produce@RRAP{*#1\href{mailto:#2}{#2}}}\frontmatter@RRAPformat}
  {}{}
}%
\makeatother
\begin{document}

\preprint{AIP/123-QED}

\title{Shearless barriers in the conservative Ikeda map}
% Force line breaks with \\
\author{Rodrigo Simile Baroni}
\affiliation{Universidade de São Paulo - USP, Instituto de Física, 05508-900, São Paulo, SP, Brasil}
\thanks{r.baroni@usp.br}

\author{Ricardo Egydio de Carvalho}
\affiliation{Universidade Estadual Paulista - UNESP, Instituto de Geociências e Ciências Exatas, Departamento de Estatística, Matemática Aplicada e Computação, 13506-900, Rio Claro, SP, Brasil}
\thanks{Senior Professor}

\author{José Danilo Szezech Junior}
\affiliation{Universidade Estadual de Ponta Grossa - UEPG, Departamento de Matemática e Estatística, 84030-900, Ponta Grossa, PR, Brasil }
\affiliation{Universidade de São Paulo - USP, Instituto de Física, 05508-900, São Paulo, SP, Brasil}

\author{Iberê Luiz Caldas}%
\affiliation{Universidade de São Paulo - USP, Instituto de Física, 05508-900, São Paulo, SP, Brasil \relax}

\date{\today}

\begin{abstract}
We investigate the dynamics of the Ikeda map in the conservative limit, where it is represented as a two-dimensional area-preserving map governed by two control parameters, $\theta$ and $\phi$. We demonstrate that the map can be interpreted as a composition of a rotation and a translation of the state vector. In the integrable case ($\phi = 0$), the map reduces to a uniform rotation by angle $\theta$ about a fixed point, independent of initial conditions. For $\phi \ne 0$, the system becomes nonintegrable, and the rotation angle acquires a coordinate dependence. The resulting rotation number profile exhibits extrema as a function of position, indicating the formation of shearless barriers. We analyze the emergence, persistence, and breakup of these barriers as the control parameters vary.
\end{abstract}

\maketitle

\section{Introduction}

The Ikeda map is a well-known system that exhibits chaotic behavior and a wide range of nonlinear dynamic phenomena. Originally proposed to describe the dynamics of light rays in a nonlinear ring cavity \cite{Ikeda1979,Ikeda1980}, its simplicity has made it a prototype model for studying dissipative nonlinear phenomena \cite{Aligood1997,Lai2011,Grebogi1987,Rech2000,Kraut2003,Kraut2003-2,FMOliveira2024}. Meanwhile, the conservative version of the system has received less attention, despite exhibiting rich dynamical behavior \cite{Kuznetsov2008}.

Conservative and dissipative systems are fundamentally different. Dissipative dynamics contract phase space volumes, leading to asymptotic states known as attractors \cite{Aligood1997,Lai2011}. On the other hand, conservative systems preserve phase volumes and often present a mixed phase space \cite{Lichtenberg1992} with the coexistence of regular (periodic and quasiperiodic) and chaotic solutions. In conservative systems, chaotic trajectories can be totally or partially bounded by regular structures that act as transport barriers \cite{MacKay1984,Viana2021}. Symplectic maps, such as the conservative Ikeda map, are the discrete-time analogue of Hamiltonian systems. In two dimensions, they can be interpreted as the Poincaré surface of section of a three-dimensional Hamiltonian flow, and preserve area and orientation \cite{Meiss1992, Lichtenberg1992}. Maps are useful because they are simpler to study than differential equations; questions regarding transport properties of phase space trajectories can be addressed with much less computational effort \cite{Meiss1992}.

Symplectic twist maps satisfy the so-called twist condition, i.e., a condition that guarantees the nondegeneracy of frequencies in the integrable limit. For such systems, in angle-action coordinates, as the action is varied, the average time increment of the angle - quantified by the rotation number - changes monotonically. The twist condition is assumed to be satisfied globally in the proofs of many results of Hamiltonian nonlinear dynamics \cite{Lichtenberg1992,Mather1982,Aubry1983,MacKay1992}, including the well-known KAM theorem. 

Nontwist systems violate the twist condition, which leads to unique phase space phenomena due to the nonmonotonic rotation number profile \cite{Szezech2009,DeCarvalho1992,Del-Castillo-Negrete1996,Morrison2000}. Of particular interest is the solution for which the rotation number is an extreme point, called the shearless curve since the derivative of the rotation number with respect to the action vanishes. These curves are robust in the sense that, as the perturbation is increased and invariant curves are destroyed, the shearless curves are roughly the last to be destroyed, limiting the transport of chaotic solutions even for strong perturbations. Evidence for shearless barriers has been reported in fluid flow experiments \cite{Behringer1991,reportDel-Castillo-Negrete1992} and in toroidal devices for plasma confinement \cite{Marcus2008,Marcus2008-2,Toufen2012}; as they have been identified in mathematical models of geophysical flows \cite{Del-Castillo-Negrete1992,Morrison2000} and magnetically confined plasmas \cite{Caldas2012-2}. More recently, it has been discovered that shearless barriers can be created even in twist systems inside resonance islands. This is often associated with the bifurcation of periodic orbits, such as tripling \cite{Dullin2000}, quadrupling \cite{Dullin2000,VieiraAbud2012}, saddle-node and pitchfork bifurcations \cite{Leal2025}.

In contrast to twist and nontwist systems, maps that violate the twist condition everywhere are globally degenerate and exhibit a constant rotation number profile. An example is the simple harmonic motion, whose angular frequency is independent of the action; many physical systems reduce to this linear oscillator to leading order. Despite this, the dynamics of globally degenerate systems under perturbations remain comparatively underexplored. To our knowledge, the best-studied representative is the web map \cite{Zaslavsky1991,Longcope1987,Murakami1988,Zaslavsky2007}, derived for a nonrelativistic particle in a uniform magnetic field subjected to a resonant electrostatic wave packet propagating perpendicular to the field. The resulting phase space consists of symmetrically arranged islands with the same rotation number, separated by a narrow chaotic layer - the stochastic web - that percolates between them. Most previous work has focused on describing particle diffusion along this web. 

Globally degenerate systems are structurally unstable \cite{Del-Castillo-Negrete1996}; nonlinear perturbation (or, in the web map, a deviation from the resonant condition) makes the system degenerate or nondegenerate. Here we study the conservative Ikeda map, whose integrable limit is globally degenerate and equivalent to a linear oscillator. We determine when a small perturbation generates a local degeneracy in the rotation number profile, numerically and with a perturbative approach. The appearance/disappearance of shearless curves, as the perturbation is increased, is linked with two bifurcations that previously were not observed in this context: a distorted pitchfork bifurcation and a subcritical period-doubling bifurcation.

The paper is organized as follows. In Section II, we present the real two‐dimensional form of the Ikeda map; the integrable limit is examined and demonstrated to be globally degenerate. The violation of the twist condition is discussed. Section III develops our numerical procedure for detecting shearless curves via rotation number extrema and explores their bifurcations for representative parameter values in association with fixed points bifurcations. Section IV explores the breakup of shearless barriers at critical values and their role as barriers to chaotic transport. Finally, Section V summarizes our findings and discusses prospects for future work. 

\section{The model}

The Ikeda map was originally proposed to describe the dynamics of a laser pulse in a nonlinear optical cavity \cite{Ikeda1979,Ikeda1980}. The system, normalized to dimensionless form \cite{Hammel1985}, is expressed in terms of a complex variable $z=x+iy$ as:
\begin{equation}\label{eq:complexIkeda}
    z_{n+1}=A+Bz_n e^{i\left(\theta-\frac{\phi}{|z_n|^2+1} \right)},
\end{equation}

\noindent where the modulus and phase of $z_n$ represent the amplitude and phase of the $n$th laser pulse exiting the cavity. The parameter $A$ represents the laser input amplitude, while $B$ is a damping parameter related to the reflection properties of cavity mirrors. The detuning of the empty cavity is given by $\theta$, and the detuning due to the nonlinear dielectric medium is described by $\phi$. Another interpretation of the Ikeda map is as an approximation to the stroboscopic map of a driven nonlinear oscillator \cite{KUZNETSOV2001,Kuznetsov2008}. 

Using Euler's formula, Eq. (\ref{eq:complexIkeda}) can be decomposed into its real and imaginary components, yielding the two-dimensional map:
\begin{equation}
\begin{aligned}\label{eq:ikeda2d}
x_{n+1} &= \Re{(z_{n+1})} = 1 + B(x_n\cos{t_n}-y_n\sin{t_n}), \\
y_{n+1} &= \Im{(z_{n+1})} = B(x_n\sin{t_n}+y_n\cos{t_n}),
\end{aligned}
\end{equation}

\noindent where we have set $A=1$ for simplicity. The angle $t_n$ is given by:
\begin{equation}{\label{eq:rotAngle}}
    t_n(x_n,y_n)=\theta-\frac{\phi}{x_n^2+y_n^2+1}.
\end{equation}

From the Jacobian matrix $J$, we find that $\det J = B^2$. If $0<B<1$, we have $\det J<1$, indicating that the map is area-contracting and dissipative. This configuration, which is the most widely studied, exhibits phenomena such as chaotic attractors, chaotic transients, crises, and other related behaviors \cite{Aligood1997,Lai2011}.

Choosing $B=1$ results in $\det J=1$, making the map area- and orientation-preserving \cite{Meiss1992}. Thus, the system is Hamiltonian, and from this point onward we will consider only this case, i.e.,
\begin{equation}
\begin{aligned}\label{eq:ikedaCons}
x_{n+1} &= 1 + x_n\cos{t_n}-y_n\sin{t_n}, \\
y_{n+1} &= x_n\sin{t_n}+y_n\cos{t_n},
\end{aligned}
\end{equation}

\noindent with $t_n$ given by Eq. (\ref{eq:rotAngle}). We note that the mapping can be interpreted as the composition of two transformations: a rotation of the state vector $[x_n,y_n]$ by the coordinate-dependent angle $t_n(x_n,y_n)$, and a translation of $x_n$ by $1$. In matrix notation, this can be written as:
\begin{equation}\label{eq:ikedaConsMatrix}
    \begin{bmatrix}
        x_{n+1} \\
        y_{n+1} 
    \end{bmatrix}
    =
    \begin{bmatrix}
        \cos{t_n} & -\sin{t_n} \\
        \sin{t_n} & \cos{t_n} 
    \end{bmatrix}
    \begin{bmatrix}
        x_{n} \\
        y_{n} 
    \end{bmatrix}
    +
    \begin{bmatrix}
        1 \\
        0 
    \end{bmatrix}.
\end{equation}

In terms of the original model, $B=1$ corresponds to the idealized case of a lossless optical cavity, i.e., the attenuation of the laser pulse after a round trip inside the cavity is neglected. Although this is unrealistic in an experimental setup, ultra-high-Q cavities have been developed in the past decades \cite{Armani2003,Lee2012,Burla2015,Wu2020}. These cavities are designed to have extremely small losses; therefore, a conservative model is a good approximation for the dynamics in time scales shorter than the damping time.

\subsection{The integrable limit}

For $\phi=0$, the map defined by Eq. (\ref{eq:ikedaConsMatrix}) reduces to:
\begin{equation}\label{eq:ikedaConsMatrixInt}
    \begin{bmatrix}
        x_{n+1} \\
        y_{n+1} 
    \end{bmatrix}
    =
    \begin{bmatrix}
        \cos{\theta} & -\sin{\theta} \\
        \sin{\theta} & \cos{\theta} 
    \end{bmatrix}
    \begin{bmatrix}
        x_{n} \\
        y_{n} 
    \end{bmatrix}
    +
    \begin{bmatrix}
        1 \\
        0 
    \end{bmatrix}.
\end{equation}

In this case, the rotation angle is simply $\theta$ and is independent of the coordinates. The fixed point of the system is found by setting $[x_{n+1},y_{n+1}]=[x_{n},y_{n}]=[x^*,y^*]$, which yields:
\begin{align}\label{eq:integrableFP}
    x^* &= \frac{1}{2}, &
    y^* &= \frac{\sin\theta}{2\,(1-\cos\theta)}.
\end{align}

Due to the translation introduced by the map, the rotations are performed around the fixed point of the map, not around the origin of the system. However, by performing a change of coordinates, the map in Eq. (\ref{eq:ikedaConsMatrixInt}) can be rewritten as a pure rotation around the origin. Let $x=u+x^*$ and $y=v+y^*$. Then, Eq. (\ref{eq:ikedaConsMatrixInt}) becomes:
\begin{equation}\label{eq:ikedaRotUV}
    \begin{bmatrix}
        u_{n+1} \\
        v_{n+1} 
    \end{bmatrix}
    =
    \begin{bmatrix}
        \cos{\theta} & -\sin{\theta} \\
        \sin{\theta} & \cos{\theta} 
    \end{bmatrix}
    \begin{bmatrix}
        u_{n} \\
        v_{n} 
    \end{bmatrix}.
\end{equation}

Equivalently, with $w=u+iv$ we have the complex map $w_{n+1}=e^{i\theta}w_n$: each iterate is a counterclockwise rotation by $\theta$ about the origin. In polar form, a point in the complex plane is represented as $w=|w|e^{i\psi}$, with $|w|=\sqrt{u^2+v^2}$ and $\psi=\arg w=\arctantwo{(v,u)}$. We wish to use $\psi$ as the angular coordinate of an action--angle canonical pair. This is achieved with the transformation:

\begin{equation}\label{eq:coordtransf}
u=\sqrt{2I}\cos\psi,\qquad
v=\sqrt{2I}\sin\psi.
\end{equation}

In terms of the new variables, Eq.~\ref{eq:ikedaRotUV} is written as:
\begin{equation}\label{eq:ikedaIntRot}
\qquad I_{n+1}=I_n,\qquad \psi_{n+1}=\psi_n+\theta \pmod{2\pi}.
\end{equation}

As expected from an integrable Hamiltonian system, the action coordinate remains unchanged under time evolution, while the angle coordinate evolves cyclically. Equation (\ref{eq:ikedaIntRot}) is the same map of a harmonic oscillator in action--angle coordinates, establishing a direct correspondence between the integrable Ikeda map and the harmonic oscillator. Our choice of action–angle variables and the associated Hamiltonian structure are detailed in Appendix~\ref{ap:action-angle}.

\subsection{The twist condition and the nonintegrable map}

A large class of area-preserving maps can be described in the form \cite{Lichtenberg1992}:
\begin{equation}
\begin{aligned}\label{eq:GeneralMap}
J_{n+1} &= J_n + \epsilon f(\vartheta_n,J_{n+1}), \\
\vartheta_{n+1} &= \vartheta_n + \Omega(J_{n+1}) + \epsilon g(\vartheta_n,J_{n+1})\quad \pmod{2\pi},
\end{aligned}
\end{equation}

\noindent with the area preservation condition $\frac{\partial f}{\partial J_{n+1}}+\frac{\partial g}{\partial \vartheta_n}=0$, and $(\vartheta,J)$ a pair of canonical variables. These maps are suitable for the study of an integrable Hamiltonian system subjected to a nonintegrable perturbation. The Hamiltonian for such a system can be written as
\begin{equation}
    H=H_0(J)+\epsilon H_1(\vartheta,J,t),
\end{equation}

\noindent where $H_0$ is the unperturbed Hamiltonian with associated angle-action variables $(\vartheta,J)$, and $H_1$ is the nonintegrable perturbation, with intensity controlled by $\epsilon$ and periodic in $\vartheta$ and $t$. The perturbing functions $f$ and $g$ are the partial derivatives of $H_1$ with respect to $\vartheta$ and $J$, respectively. The function $\Omega$ is associated with the frequency of the unperturbed Hamiltonian, $\frac{\partial H_0(J)}{\partial J}$. The unperturbed system's solutions are confined to invariant tori, and are either periodic (rational $\Omega$) or quasiperiodic (irrational $\Omega$).

The nondegeneracy condition for the frequencies of a Hamiltonian flux is $\frac{\partial^2H_0}{\partial J^2}\ne0$, which corresponds to the twist condition for maps: $\frac{\partial \vartheta_{n+1}}{\partial J_n}\ne0$. A system that satisfies the twist condition is referred to as a twist system, and this property is required for the rigorous proof of many results in Hamiltonian dynamics \cite{Lichtenberg1992,Mather1982,Aubry1983,MacKay1992}, such as the conventional KAM theorem and the Poincaré-Birkhoff theorem. These theorems provide a general qualitative description of the effects of a small nonintegrable perturbation: most irrational invariant tori persist, albeit slightly deformed, and, around the rational tori, chains of alternating elliptic islands and hyperbolic saddle points are created. Chaotic layers are formed in the vicinity of the saddle points.

The integrable map described by Eq. (\ref{eq:ikedaIntRot}) violates the twist condition globally, as $\frac{\partial \psi_{n+1}}{\partial I_n}=0$ for all $I_n$. Another example of a globally degenerate system is the web map \cite{Zaslavsky1991,Longcope1987,Murakami1988,Zaslavsky2007}, that models the motion of a charged particle in a constant magnetic field, and subject to an electrostatic wave packet propagating perpendicularly to the magnetic field. The Hamiltonian of the system is that of a periodically kicked linear oscillator under a resonant condition. The main property of the web map is that its phase space exhibits an unbounded domain of chaos that percolates around symmetrically distributed islands. 

It is known that globally degenerate Hamiltonian systems are not structurally stable \cite{Del-Castillo-Negrete1996}; the presence of a nonlinear perturbation and a small deviation from the resonant condition can make the system nondegenerate or locally degenerate. Of particular interest is the case with local degeneracies, known as nontwist systems. The curve along which the twist condition is violated is referred to as the shearless curve. This curve acts as a robust transport barrier, as it is highly resistant to perturbations, being roughly the last one to be destroyed, inhibiting chaotic transport between phase space regions \cite{Viana2021}. Even after its destruction, sticky behavior is often observed where the shearless curve once was, resulting in a partial transport barrier. Moreover, small changes in the control parameters may cause the shearless curve to reappear \cite{Szezech2009,Mugnaine2024}. 

Another effect of the local violation of the twist condition is the formation of twin resonances on each side of the shearless curve. The separatrices of these resonances experience a reconnection process near the shearless curve, a bifurcation that alters the phase space topology in that region \cite{DeCarvalho1992,Wurm2005}. During the reconnection process, meandering tori may appear \cite{Wurm2005}. 

The shearless curve can be found by calculating the rotation number profile for a set of initial conditions and checking whether it has an extreme value \cite{VieiraAbud2012,Osorio-Quiroga2024,Leal2025}. The rotation number $\omega(\vartheta_0,J_0)$ corresponds to the average angular displacement of the orbit generated from the initial condition $\mathbf{X}_0=(\vartheta_0,J_0)$. If $\omega$ is a rational number $r/s$, the orbit is periodic with period $s$; if it is irrational, the orbit is quasiperiodic. To compute $\omega$, we use the superconvergent method presented in Refs. \cite{Das2017,Das2018,Sales2022,Osorio-Quiroga2024,Tong2024}, which provides a faster convergence compared to the standard average. It is defined as:
\begin{equation}
    \omega(\mathbf{z}_n)=\frac{1}{2\pi} \sum_{n=0}^{N-1} \hat{w}_{n,N}[ \Pi (M(\mathbf{X}_n)) - \Pi(\mathbf{X}_n)],
\end{equation}
\begin{equation}
    \hat{w}_{n,N} = \frac{w(n/N)}{\sum_{n=0}^{N-1} w(n/N)},
\end{equation}
\begin{equation}
    w(\xi) = \begin{cases}
\exp\left( \frac{-1}{\xi(1-\xi)} \right) & \text{for } \xi \in (0,1), \\
0 & \text{for } \xi \notin (0,1),
\end{cases}
\end{equation}

\noindent where $M(\mathbf{X}_n)$ is the map applied to $\mathbf{X}_n$ and $\Pi$ is a suitable angular projection. In maps of the form of Eq. (\ref{eq:GeneralMap}), where the dynamical variable $\vartheta_n$ is the angle coordinate itself, we have $\Pi(\mathbf{X}_n)=\vartheta_n$. For the conservative Ikeda map, as given by Eq. (\ref{eq:ikedaCons}), we consider rotations around a fixed point of the map, so $\Pi(\mathbf{z}_n)=\psi_n=\arctantwo{(y_n-y^*,x_n-x^*)}$. 
The initial conditions considered to calculate the rotation number will be distributed along the line $x=1/2$, as this is a symmetry line of the system (see Appendix \ref{ap:symmetry}). 

Considering the nonintegrable map from Eq. (\ref{eq:ikedaCons}), it is not straightforward to perform coordinate changes to transform it into an action-angle form, as we did for the integrable map in Eq. (\ref{eq:ikedaConsMatrixInt}), which led to Eq. (\ref{eq:ikedaIntRot}), and explicitly revealed the global degeneracy of the frequencies. This difficulty arises because the fixed points of the nonintegrable map must be found numerically, and due to the coordinate-dependent angle $t_n$ given by Eq. (\ref{eq:rotAngle}). Thus, to examine how the global degeneracy of the rotation number profile is affected by a small perturbation, we carry out a numerical experiment and a perturbative approach. The perturbative results are presented in the Appendix \ref{ap:Perturbative}, where we derive an analytical expression for the rotation number for $\phi<<1$ (Eq. (\ref{eq:rotNumAnl})), and conclude that the profile is nonmonotonic for $\theta\in(0,\pi/2)\cup(3\pi/2,2\pi)$.

\begin{figure}
\centering
    \begin{minipage}{0.8\linewidth}
        \centering
        \includegraphics[width=\textwidth]{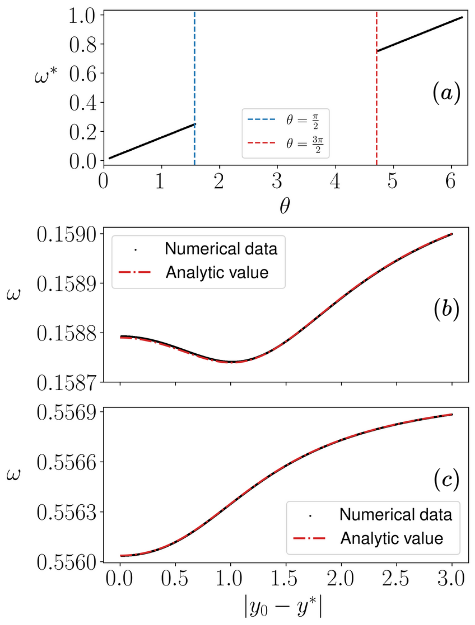}
    \end{minipage}
\caption{\label{fig:fig1} For $\phi=0.01$, (a) extreme value of the rotation number  as a function of $\theta$. In the interval $\theta\in\left(\pi/2,3\pi/4\right)$, the rotation number profile is monotonic. (b) Nonmonotonic rotation number profile obtained for $\theta=1$. (c) Monotonic rotation number profile obtained for $\theta=3.5$.}
\end{figure}

In the numerical experiment we set $\phi=0.01$, a small perturbation value, and check if the rotation number profile exhibits an extreme point. The diagram in Fig. \ref{fig:fig1}(a) shows how the extreme value of the rotation number, $\omega^*$, changes with $\theta$ for $\phi=0.01$. We observe that $\omega^*$ increases linearly with $\theta$, with the absence of extreme values in the interval $\theta\in\left(\pi/2,3\pi/4\right)$ justified by the perturbative approach. Panels (b) and (c) show the analytic and numerical rotation number profile, for $\theta=1$ (which is nonmonotonic) and $\theta=3.5$ (which is monotonic), respectively. As will be shown in the next section, it is possible that, by increasing $\phi$, a nonmonotonic rotation number profile could emerge even for $\theta$ values in the interval where no extreme points were observed for $\phi\ll1$.

\section{Shearless curve bifurcations}

To investigate the presence of the shearless curve as the system parameters vary, we use the following method to construct bifurcation diagrams of the shearless curve. For a fixed value of $\theta$, we consider values of $\phi$ uniformly distributed in the interval $[0,2\pi]$. For each parameter combination, we compute the rotation number profile along a vertical line segment of initial conditions, extending from the fixed point to an appropriately chosen minimum value of $y_0$. We then search for extreme points in the rotation number profile by identifying values of $y_0$ for which $d\omega/dy_0=0$. If such a point exists, $(x^*,y_0)$ corresponds to a point on a shearless curve. We compare the resulting diagram with the bifurcation diagram of the fixed points, as we have observed that the creation or destruction of the shearless curve is often associated with bifurcations of the fixed points.

\subsection{Case $\theta=1$}

\begin{figure}[h]
\centering
    \begin{minipage}{0.8\linewidth}
        \centering
        \includegraphics[width=\textwidth]{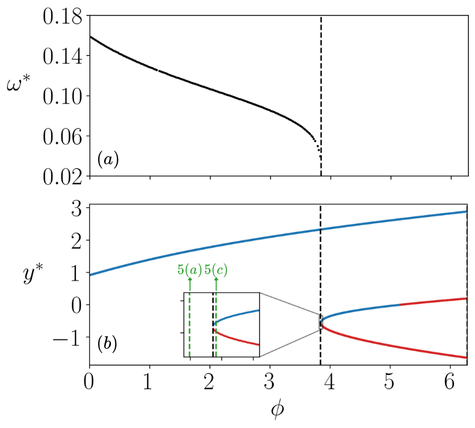}
    \end{minipage}
\caption{\label{fig:fig2} For $\theta=1$, (a) bifurcation diagram of the shearless curve and (b) bifurcation diagram of the fixed points, where blue denotes stability and red denotes instability. The saddle-center bifurcation around $\phi=3.8$ (marked by the vertical black line) is associated with the disappearance of the shearless curve. The vertical green lines in the inset indicate the two consecutive $\phi$ values considered in Fig. \ref{fig:fig3}.}
\end{figure}

We begin by analyzing the case $\theta=1$. Figure \ref{fig:fig2}(a) shows the bifurcation diagram of the shearless curve, where we observe that the extreme rotation number $\omega^*$ decreases as $\phi$ increases. Around $\phi=3.8$, the shearless curve disappears, coinciding with a saddle-center bifurcation that creates a pair of stable and unstable fixed points. This type of bifurcation, where a second stable fixed point arises abruptly along an unstable one, is known as a distorted pitchfork bifurcation \cite{Tel2006} or imperfect bifurcation \cite{Strogatz2018}, and is associated with some asymmetry in the system. The bifurcation diagram of the fixed points is shown in Fig. \ref{fig:fig2}(b), where the $y^*$ coordinate of the fixed points is plotted. Stability is indicated by color: blue represents stable fixed points, and red represents unstable fixed points. The rotations for the rotation number profile are always computed with the fixed point with the higher $y^*$ value, which corresponds to the continuation of the integrable map's fixed point.

\begin{figure}
\centering
    \begin{minipage}{1\linewidth}
        \centering
        \includegraphics[width=\textwidth]{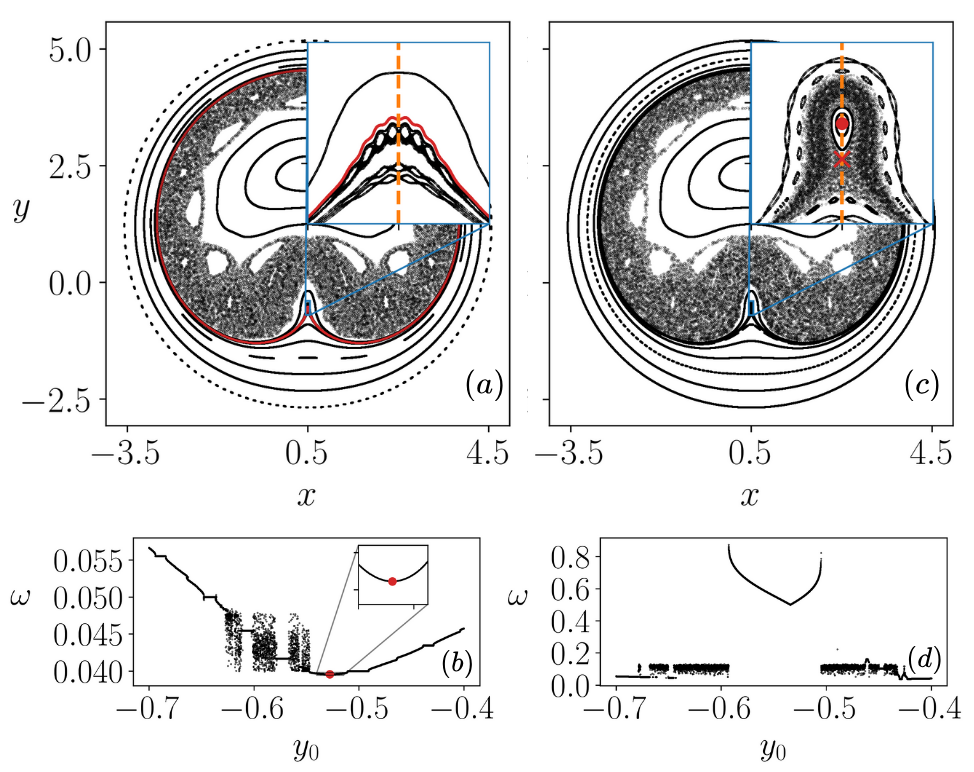}
    \end{minipage}
\caption{\label{fig:fig3} Phase space and rotation number profile for $\theta=1$: (a) and (b) show $\phi=3.8245$, before the saddle-center bifurcation, and (c) and (d) show $\phi=3.8456$, after the saddle-center bifurcation.}
\end{figure}

Figure \ref{fig:fig3}(a) shows the phase space for $\theta=1$ and $\phi=3.8245$, just before the saddle-center bifurcation. The shearless curve is shown in red, and the inset provides an amplified view of the region where the bifurcation will occur. The dashed line represents the initial conditions used to compute the rotation number profile in panel (b), where the extreme point is marked in red. Panel (c) shows the phase space for $\phi=3.8456$, after the saddle-center bifurcation has occurred. The inset shows the fixed points created by the bifurcation: the stable fixed point is marked as a circle, and the unstable one as a cross. The rotation number profile in panel (d) is computed using the same initial conditions as in panel (b), but no extreme points are found; the observed nondifferentiable local minimum corresponds to the elliptic fixed point created by the bifurcation \cite{Wurm2005}.

\begin{figure}
\centering
    \begin{minipage}{1\linewidth}
        \centering
        \includegraphics[width=\textwidth]{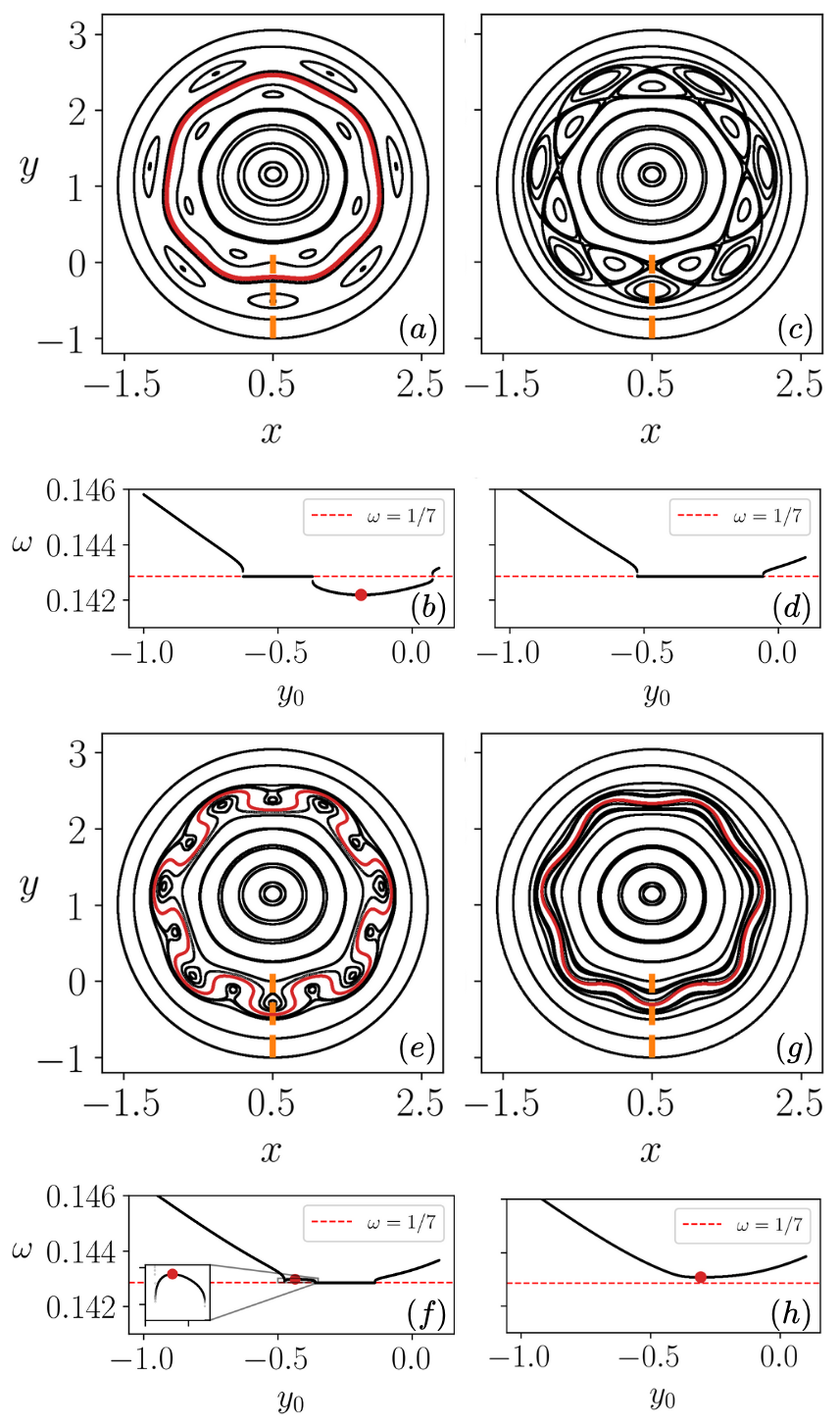}%\\(a)
    \end{minipage}
\caption{\label{fig:fig4} Reconnection-collision sequence of the $\omega=1/7$ twin resonances. Phase spaces (upper row) and rotation number profiles (lower row) for $\theta=1$ and decreasing values of $\phi$; the dashed line segment at $x=0.5$ in the phase spaces represents the initial conditions used to compute the rotation number profile. In (a) and (b), $\phi=0.4833$; (c) and (d), $\phi=0.46745$; (e) and (f), $\phi=0.4623$; (g) and (h), $\phi=0.455$.}
\end{figure}

Figure \ref{fig:fig4} illustrates the reconnection-collision sequence of the $\omega=1/7$ twin resonances, as well as the  meandering tori. Starting with decreasing values of $\phi$, panel (a) shows the shearless curve in red, identified by the global minimum in the rotation number profile in panel (b), and the twin period-7 resonances. One resonance lies inside the region delimited by the shearless curve, while the other lies outside of it. In panel (c), $\phi$ is decreased to the reconnection threshold, and the corresponding rotation number profile in panel (d) shows no extreme points, with the global minima corresponding to the $\omega=1/7$ plateau. In panel (e), after the reconnection process, the interior and exterior of the shearless curve region now contain elliptic and hyperbolic points from the opposing resonances. In panel (f), we observe a local maximum that corresponds to the shearless curve, and is associated with meandering tori. In panel (g), the elliptic and hyperbolic points of the opposing resonances have collided and mutually annihilated. Finally, in panel (h), the rotation number profile shows a global minimum corresponding to the shearless curve.

\subsection{Case $\theta=3.5$}

\begin{figure}[h]
\centering
    \begin{minipage}{0.8\linewidth}
        \centering
        \includegraphics[width=\textwidth]{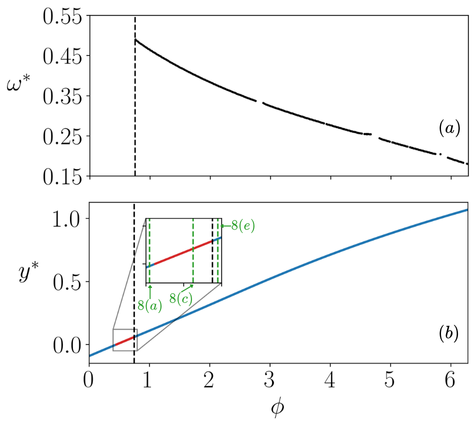}
    \end{minipage}
\caption{\label{fig:fig5} For $\theta=3.5$, (a) bifurcation diagram of the shearless curve and (b) bifurcation diagram of the fixed point, where blue denotes stability and red denotes instability. A subcritical period-doubling bifurcation around $\phi=0.7$ (marked by the vertical black line) is associated with the appearance of the shearless curve. The vertical green lines in the inset indicate the three consecutive $\phi$ values considered in Fig. \ref{fig:fig6}.}
\end{figure}

The shearless curve bifurcation diagram for $\theta=3.5$ is shown in Figure \ref{fig:fig5}(a). In agreement with the result in Fig. \ref{fig:fig1}(a), no extreme point is found in the rotation number profile for small values of $\phi$. Around $\phi=1$, a shearless curve emerges, and the extreme rotation number decreases as $\phi$ increases. The gaps in the diagram correspond either to reconnections of twin resonances, during which the shearless curve disappears, or the actual breakup and reappearance of the shearless curve, illustrating sensitivity to the control parameters. Panel (b) shows the stability bifurcation of the fixed point. A single fixed point exists for this parameter range, but its stability switches from stable to unstable through a period doubling bifurcation. Upon further increasing $\phi$, the fixed point becomes stable again, coinciding with the creation of the shearless curve.  

\begin{figure}
\centering
    \begin{minipage}{1\linewidth}
        \centering
        \includegraphics[width=\textwidth]{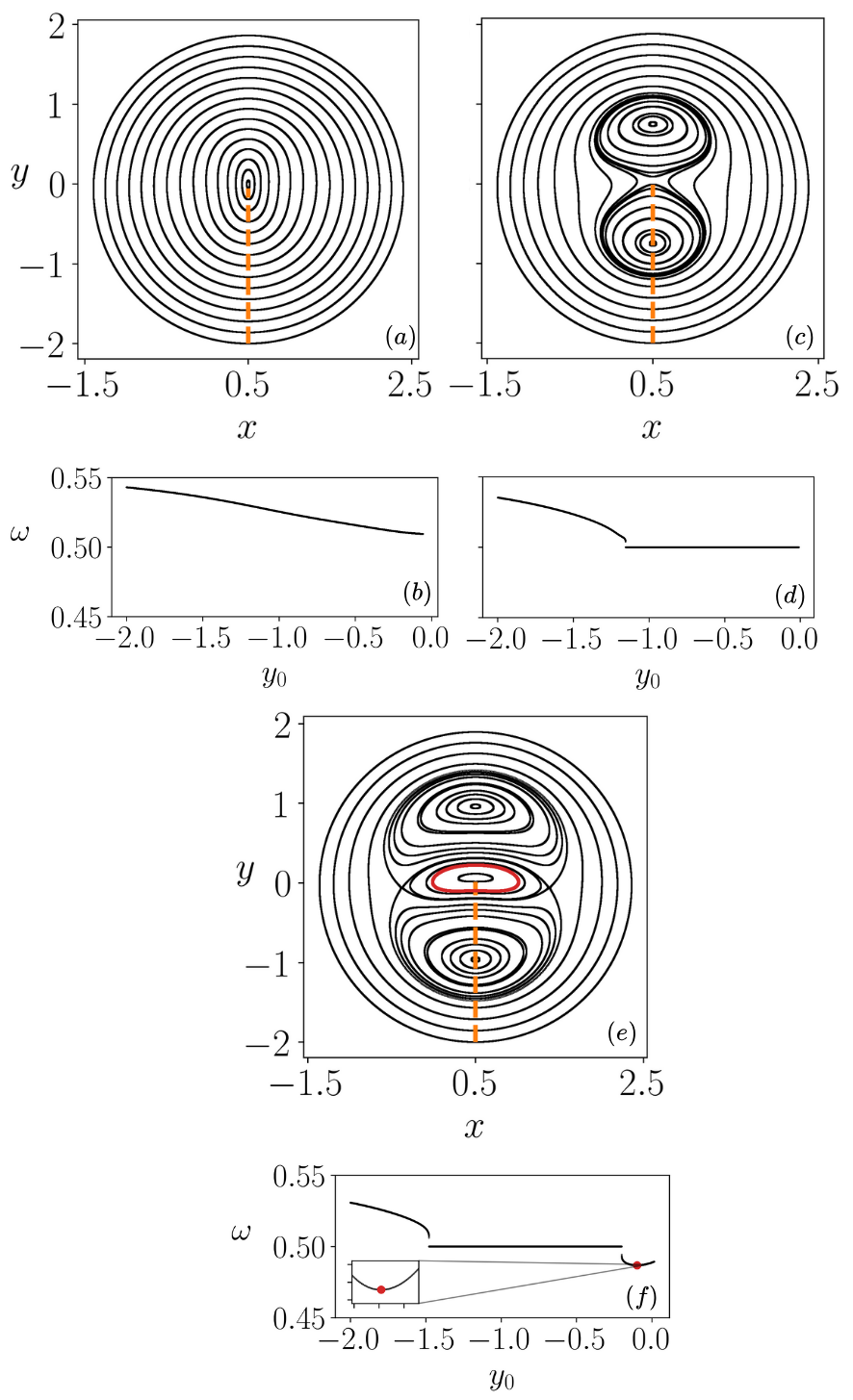}
    \end{minipage}
    
\caption{\label{fig:fig6} Phase space and rotation number profile for $\theta=3.5$: (a) and (b) show $\phi=0.42$, before the supercritical period-doubling bifurcation; (c) and (d) show $\phi=0.65$, after the supercritical period-doubling bifurcation and before the subcritical period-doubling bifurcation; and (e) and (f) show $\phi=0.78$, after the subcritical period-doubling bifurcation, which also creates a shearless curve.}
\end{figure}

Figure \ref{fig:fig6} shows the phase spaces (upper row) and rotation number profiles (lower row), illustrating the bifurcations that lead to the creation of the shearless curve. In panels (a) and (b), we consider $\phi=0.42$, before the period-doubling bifurcation of the fixed point, where the rotation number profile is monotonic. In panels (c) and (d) we have $\phi=0.65$, after the supercritical period-doubling bifurcation and before the subcritical period-doubling bifurcation, the fixed point is unstable and a period-2 resonance is observed. In panels (e) and (f) we consider $\phi=0.78$, after the subcritical period-doubling bifurcation, the rotation number profile shows a minimum point, which identifies the shearless curve, shown in red in panel (e).

\section{Shearless curve breakup}

To illustrate both the breakup of the shearless curve as the perturbation is increased and its role as a barrier to chaotic transport, we search in the parameter space for the values that result in a shearless curve with a specific frequency. The numerical scheme is as follows: suppose we are investigating the breakup of the shearless curve with $\omega^*=\Omega$. Starting with $\phi_0=0$, we know from the integrable map of Eq. (\ref{eq:ikedaIntRot}) that the rotation number profile is globally degenerate with $\omega^*=\theta/2\pi$. We then choose $\theta_0=2\pi\Omega$ and consider $\phi_1=\phi_0+\delta$. Next, using the bisection method, we search for a value $\theta_1\in[\theta_0-\delta,\theta_0+\delta]$ such that the rotation number profile for $\theta_1$ has the extreme value $\omega^*=\Omega$. This procedure is repeated for increasing values of $\phi$ until the shearless curve with frequency $\Omega$ is no longer found, indicating that it has broken.

\begin{figure}
\centering
    \begin{minipage}{0.8\linewidth}
        \centering
        \includegraphics[width=\textwidth]{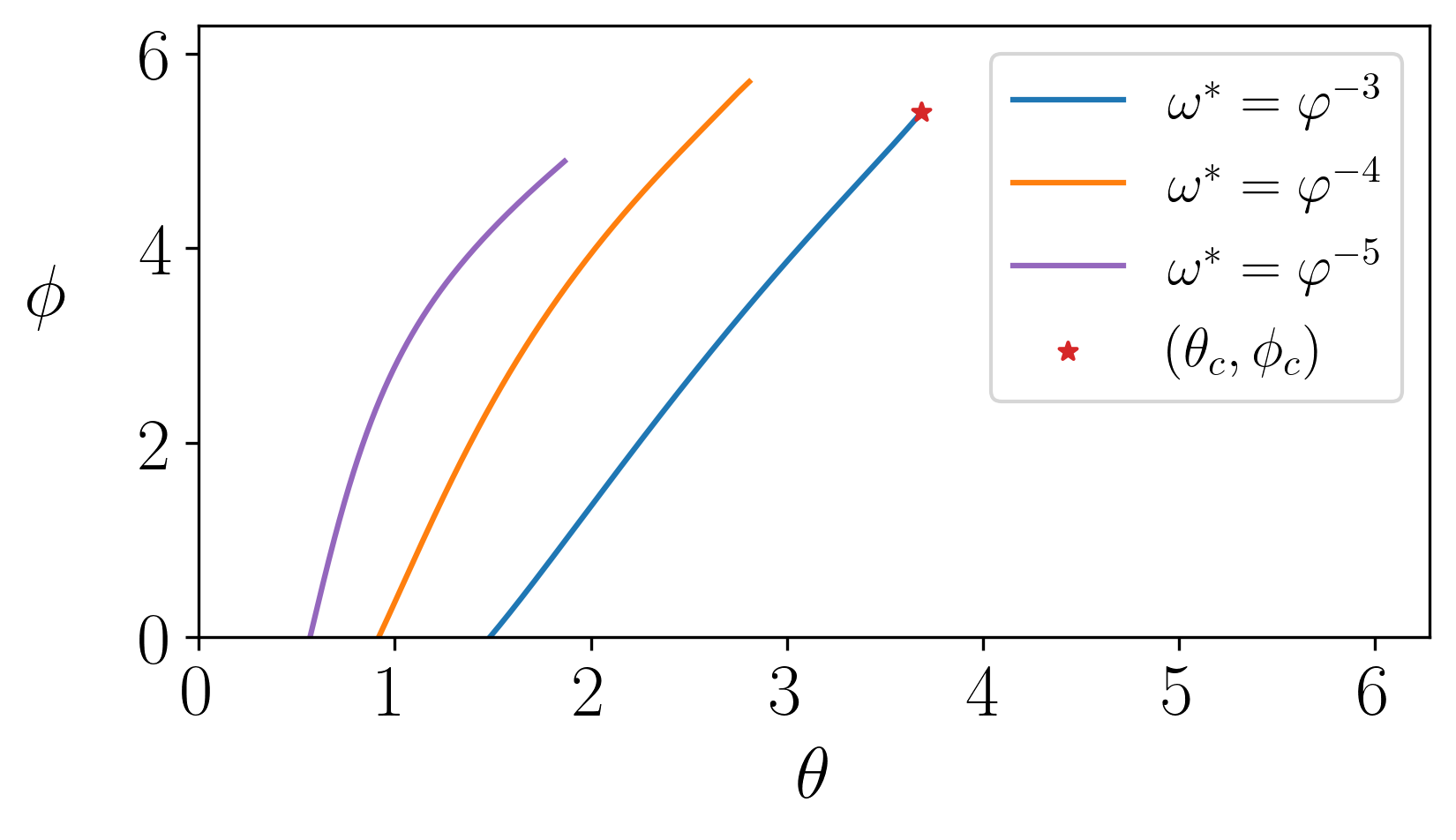}
    \end{minipage}
\caption{\label{fig:fig7} Parameter space of the conservative Ikeda map, with curves representing the parameters for which a shearless curve with a specific frequency was observed. The red star marks the critical parameter $(\theta,\phi)=(\theta_c,\phi_c)$ for which the $\omega=\varphi^{-3}$ shearless curve is in the imminence of breakup; the corresponding phase space is shown in Fig. \ref{fig:fig8}(a).
}
\end{figure}

As shown in Fig. \ref{fig:fig1}, not all values of $\theta$ will generate a locally degenerate rotation number profile when perturbed by $\phi$. Therefore, the frequency $\Omega$ of the shearless curves that can be found with this method is restricted to those within the ranges $0<2\pi\Omega<\pi/2$ and $3\pi/2<2\pi\Omega<2\pi$. We apply this method for negative powers of the golden ratio, $\varphi=\frac{1+\sqrt{5}}{2}$, that satisfy this restriction: $\varphi^{-3}\approx0.236$, $\varphi^{-4}\approx0.146$ and $\varphi^{-5}\approx0.09$. The results are shown in the parameter space of Figure \ref{fig:fig7}, where the blue, orange and purple curves represent $\omega^*=\varphi^{-3}$, $\omega^*=\varphi^{-4}$ and $\omega^*=\varphi^{-5}$, respectively. 

\begin{figure}
\centering
    \begin{minipage}{0.8\linewidth}
        \centering
        \includegraphics[width=\textwidth]{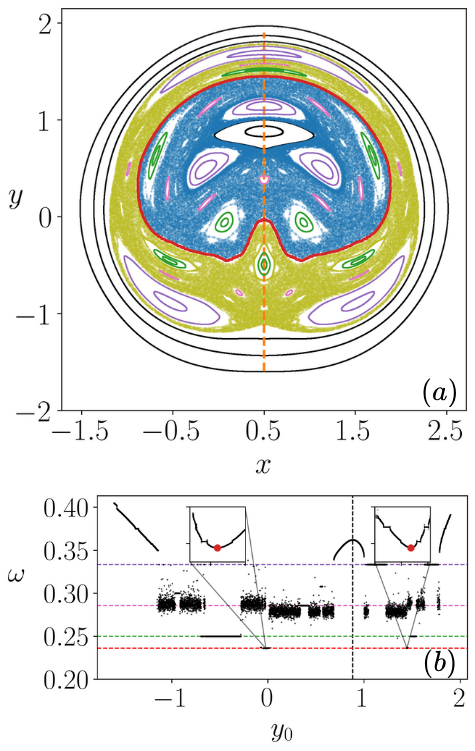}
    \end{minipage}
\caption{\label{fig:fig8} Phase space and rotation number profile for $(\theta,\phi)=(\theta_c,\phi_c)$, where the $\omega=\varphi^{-3}$ shearless curve is close to breaking up. In panel (a) the phase space is shown with the shearless curve in red, and some twin resonances are highlighted:  $\omega=1/3$ in purple, $\omega=1/4$ in green, and $\omega=2/7$ in pink. Panel (b) shows the rotation number profile computed along the dashed line in (a); the colored horizontal dashed lines correspond to the rotation number of the highlighted twin resonances and coincide with their plateaus. The insets show the extreme values of the rotation number profile.}
\end{figure}

Figure \ref{fig:fig8} illustrates the $\omega=\varphi^{-3}$ shearless curve near the point of break-up, for $(\theta,\phi)=(\theta_c,\phi_c)=(3.7008749277131261, 5.4248044343970880)$. Panel (a) shows the phase space in the map's original coordinates, $(x,y)$, with the shearless curve (in red) preventing transport between the two chaotic seas. Some twin resonances are highlighted: $\omega=1/3$ in purple, $\omega=1/4$ in green, and $\omega=2/7$ in pink. Two initial conditions in the chaotic regions are iterated $n=10^4$ times; one, colored blue, is considered within the region delimited by the shearless curve, and the second one, colored olive, is outside the region delimited by the shearless curve. We observe that the shearless curve acts as a barrier to chaotic transport, as there is no mixing between those regions. The rotation number profile, computed over the dashed line in panel (a), is shown in panel (b). The $\omega$ values of the highlighted resonances in the phase space are represented by the dashed horizontal lines, colored accordingly; coinciding with the corresponding plateaus. The red horizontal dashed line represents $\omega=\varphi^{-3}$, and the insets show amplifications of the minima of the profile, marked in red, which indicate the shearless curve. Although two minima are observed, they correspond to the same shearless curve. From the phase space in panel (a), it can be seen that the line of initial conditions used for the rotation number profile crosses the fixed point $(x^*,y^*)$, around which the rotations are computed, so some invariant curves are considered twice. The black vertical dashed line in panel (b) is at $y_0=y^*$.  

\begin{figure}
\centering
    \begin{minipage}{1\linewidth}
        \centering
        \includegraphics[width=\textwidth]{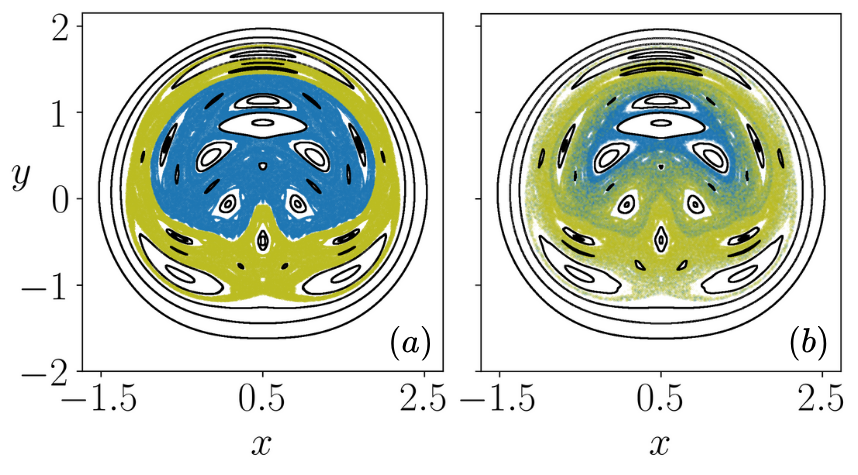}
    \end{minipage}
\caption{\label{fig:fig9} Phase spaces after the breakup of the shearless curve. In both cases two initial conditions in the chaotic sea are iterated $10^4$ times, and colored blue and olive. In panel (a), for $(\theta,\phi)=(\theta_c+10^{-2},\phi_c+10^{-2})$, though the shearless curve has been destroyed, the chaotic regions did not mix for the number of iterations considered. In panel (b), for $(\theta,\phi)=(\theta_c+2\times10^{-2},\phi_c+2\times10^{-2})$, the chaotic seas mix with the number of iterations considered.}
\end{figure}

Figure \ref{fig:fig9}(a) shows the phase space for $(\theta,\phi)=(\theta_c+10^{-2},\phi_c+10^{-2})$, past the break up of the shearless curve. Even though the shearless curve was broken, no mixing of the chaotic regions is observed with $10^4$ iterations. However, iterating further, the trajectories leave the apparently delimited area; we verified mixing with $10^5$ iterations. This is because even after the break up of the shearless curve, the transport is observed to be reduced in the region it used to be. In panel (b), we considered $(\theta,\phi)=(\theta_c+2\times10^{-2},\phi_c+2\times10^{-2})$, and $10^4$ iterations of the chaotic orbits were enough to observe the mixing of the regions. 

Interpreting the results of this section in terms of the ideal optical cavity, the shearless curve acts as a barrier that bounds the chaotic fluctuations of the field’s amplitude and phase. For parameters matching Fig. \ref{fig:fig8} and an initial condition in the olive chaotic sea, the pulse amplitude and phase would evolve chaotically but remain confined to that sea, never crossing into the blue chaotic sea; the converse holds for initial conditions on the other side. After the shearless curve breaks (Fig. \ref{fig:fig9}), mixing of the two amplitude-phase regimes becomes possible, though typically only after long transients.

% \newpage

\section{Conclusions}

In this work, we investigated the formation, persistence, and breakup of shearless barriers in the conservative Ikeda map, a two-dimensional area-preserving system governed by parameters $\theta$ and $\phi$. Starting from the integrable limit ($\phi = 0$), we showed that the map reduces to a uniform rotation about a fixed point and is globally degenerate, violating the twist condition everywhere. Introducing the perturbative parameter $\phi$ breaks this global degeneracy, with some $\theta$ yielding nonmonotonic locally degenerate rotation number profiles whose extrema mark the emergence of the shearless curves. An analytical expression for the rotation number was obtained with perturbation theory, and it was shown that, for small $\phi$, the profile is nonmonotonic if $\cos\theta>0$.

Key findings include the association of shearless barriers with bifurcations of fixed points, such as saddle-center and period-doubling bifurcations. For $\theta = 1$, the disappearance of the shearless curve coincides with a distorted pitchfork bifurcation, while for $\theta = 3.5$, the emergence of a shearless barrier follows a subcritical period-doubling bifurcation. The reconnection-collision sequences of twin resonances and the formation of meandering tori further illustrate the intricate phase space dynamics near shearless curves.

We demonstrated that shearless barriers act as effective transport inhibitors, confining chaotic trajectories even under strong perturbations. Their breakup at critical parameter values allows chaotic mixing, though remnants of these barriers can delay transport. Numerical tracking of shearless curves with frequencies related to the golden ratio highlighted their resilience and parameter-dependent thresholds for destruction.

These results contribute to the broader understanding of transport phenomena in Hamiltonian systems, as well as present a novel approach to the well-known Ikeda map. Future studies could further explore the discrete symmetries of the map and the mixing of chaotic trajectories. The methodology developed here, combining rotation number analysis and bifurcation tracking, provides a versatile framework for studying transport barriers in diverse conservative systems.

\appendix

\

\section{Action-angle coordinates for the integrable map}\label{ap:action-angle}

The conservative and integrable Ikeda map in $(u,v)$ variables (Eq. (\ref{eq:ikedaConsMatrixInt})) can be derived from the flow generated by the Hamiltonian $H=-\frac{\omega_0}{2}(u^2+v^2)$, with $u$ the generalized coordinate and $v$ the conjugate momentum. Note that this Hamiltonian differs from the harmonic oscillator by a minus sign; this is because the harmonic oscillator describes clockwise rotations in the $(u,v)$ plane, while, as we have shown, the integrable Ikeda map describes counterclockwise rotations. 

We adopt the action-angle transformation $(u,v)\mapsto (\psi,I)$ from Eq.~(\ref{eq:coordtransf}), taking $\psi=\arg(w)$ with $w=u+iv$ as the angle, since it measures rotation in the complex plane. The symplectic form in $(u,v)$ is:
\begin{equation}
    \bar\omega = du\wedge dv,
\end{equation}
with
\begin{align}\label{eq:differentials}
    du&=-\sqrt{2I}\,\sin\psi\,d\psi+\frac{\cos\psi}{\sqrt{2I}}\,dI,\\
    dv&=\;\;\sqrt{2I}\,\cos\psi\,d\psi+\frac{\sin\psi}{\sqrt{2I}}\,dI.
\end{align}
A direct computation gives:
\begin{equation}\label{eq:symForm}
    \bar\omega
    = du\wedge dv
    = -\,d\psi\wedge dI.
\end{equation}

Equation (\ref{eq:symForm}) shows that $(u,v)\mapsto (\psi,I)$ reverses the orientation of the symplectic form, so the proper ordering of the variables is $(I,\psi)$. Thus, $I$ plays the role of generalized position and $\psi$ that of the conjugate momentum. The Hamiltonian is $H=-\omega_0I$, and the equations of motion are:

\begin{equation}\label{eq:EOMIpsi}
    \dot{I}=\frac{\partial H}{\partial\psi}=0, \qquad
    \dot{\psi}=-\frac{\partial H}{\partial I}=\omega_0.
\end{equation}

The negative sign in the Hamiltonian is compensated by the sign in Hamilton's equation, so the dynamics agree with the harmonic oscillator in action-angle variables.

\section{Symmetry properties of the map}\label{ap:symmetry}

Discrete symmetries of Hamiltonian systems are useful to understand the organization of periodic orbits and to find those orbits. A detailed discussion of symmetries in dynamical systems can be found in Ref. \cite{Lamb1998}, and in the context of nontwist systems in Refs. \cite{Petrisor2001, SHINOHARA1998}. 
A transformation $T$ is called a symmetry of a map $M$ if $M=T^{-1}MT$. A transformation $I_1$ is called a time-reversal symmetry of $M$ if $M^{-1}=I_1^{-1}MI_1$. Furthermore, $I_1$ is an involution if it is its own inverse, that is, $I_1^2=\text{Id}$. In this case, it can be used to construct a second time reversal symmetry of $M$, $I_2=MI_0$, which is also an involution, and the map can be decomposed as $M=I_2I_1$. Maps that obey this property are called reversible.

The conservative Ikeda map has the following time reversal symmetry:

\begin{equation}\label{eq:involution_1}
    I_1(z)=1-\bar{z},
\end{equation}

\noindent where $\bar{z}$ is the complex conjugate of $z$. Equation (\ref{eq:involution_1}) represents a reflection along the line $x=1/2$ and is also an involution. It can be used to construct a second involution from $MI_1$:

\begin{equation}
    I_2(z)=1+(1-\bar{z})e^{it(I_1(z))}.
\end{equation}

The symmetry lines of the map are the set of fixed points of the involutions $\Gamma_i=\{z|I_i(z)=z\}$. Symmetry lines are useful to the numerical search for symmetric periodic orbits, i.e., periodic orbits with a point belonging to the symmetry line. Considering $I_1$, we find: 

\begin{equation}
    \Gamma_1=\biggl\{(x,y)|x=\frac{1}{2}\biggl\},
\end{equation}

\noindent while $I_2$ furnishes

\begin{equation}\label{eq:SymLine2}
    \Gamma_2=\biggl\{(x,y)|y=\frac{(1-x)\sin{(t(I_1(z))}}{1-\cos{(t(I_1(z))}}\biggr\}.
\end{equation}

Equation (\ref{eq:SymLine2}) needs to be solved numerically for the nonintegrable map. 

\

\section{Perturbative approach}\label{ap:Perturbative}

The goal of this section is to obtain an analytical expression for the rotation number of the conservative Ikeda map through perturbation theory. Choosing $A=B=1$ in Eq. (\ref{eq:complexIkeda}), we obtain the map:

\begin{equation}\label{eq:fullMap}
    z_{n+1} = F(z_n; \theta, \phi)=1+z_n e^{i\left(\theta-\frac{\phi}{|z_n|^2+1} \right)}.
\end{equation}

Considering $\phi\ll1$ and expanding up to first order, we find:

\begin{equation}\label{eq:expandMap}
    z_{n+1} = 1+z_n e^{i\theta}-iz_n e^{i\theta}\frac{\phi}{|z_n|^2+1} + \mathcal{O}(\phi^2).
\end{equation}

\noindent which is a quasi-integrable form: the first two terms in the right-hand side describe the integrable part of the system, and the term proportional to $\phi$ is the perturbation. As the rotation number is computed from the rotations of a state vector around the fixed point $z^*$, we look for an expression in the form:

\begin{equation}\label{eq:FPexpansion}
    z^*(\phi) = z_0^* + \phi z_1^* + \mathcal{O}(\phi^2)
\end{equation}

\noindent where \( z_0, z_1 \in \mathbb{C} \) are coefficients to be determined. The fixed point condition is $z^*(\phi) = F(z^*(\phi); \theta, \phi)$, and the coefficients of Eq. \ref{eq:FPexpansion} are determined by expanding both sides and matching the coefficients. We find:

\begin{equation}
    z_0^*=\frac{1}{1-e^{i\theta}},
\end{equation}

\noindent which is the complex form of the integrable map's fixed point (Eq. \ref{eq:integrableFP}), and the first-order correction is:

\begin{equation}
    z_1^*=\frac{i}{3-2\cos{\theta}}.
\end{equation}

To analyze the rotations around the fixed point, we write an initial point as $z_0=z^*+|w_0|e^{i\psi_0}$, where $|w_0|=\sqrt{\Re{(z_0-z^*)}^2+\Im{(z_0-z^*)}^2}$ is the initial distance from the fixed point and $\psi_0$ the initial phase. Its first iterate can be written as $z_1=z^*+|w_1|e^{i(\psi_0+\Theta)}$, where the angular increment after a single iteration, $\Theta$, can be expressed as:

\begin{align}
    \Theta(|w_0|,\psi_0;\theta,\phi)&=\arg(z_1-z^*)-\arg(z_0-z^*),\\
    &=\arg(\delta_z)-\psi_0.
\end{align}

We look for the perturbative expansion:

\begin{equation}\label{eq:alphaExpansion}
    \Theta(|w_0|,\psi_0;\theta,\phi)=\Theta_0+\phi\Theta_1+\mathcal{O}(\phi^2).
\end{equation}

To determine the coefficients \( \Theta_0, \Theta_1 \), we first expand:

\begin{equation}\label{eq:deltazExpansion}
    \delta z=z_1-z^*=a_0+\phi a_1+\mathcal{O}(\phi^2),
\end{equation}

\noindent where $z_1=F(z_0; \theta, \phi)$. Then, the coefficients of Eq.~(\ref{eq:alphaExpansion}) are obtained from \footnote{Eq.~(\ref{eq:argExpansion}) is obtained from the complex log definition: $\log(\delta z)=\log|\delta z|+i\arg (\delta z)$. Then, substituting in Eq.~(\ref{eq:deltazExpansion}) in the LHS, expanding $\log(1+\epsilon)\approx\epsilon-\frac{1}{2}\epsilon^2$, and taking the imaginary part of the resulting expression yields Eq. (\ref{eq:argExpansion}).}:

\begin{equation}\label{eq:argExpansion}
    \arg(\delta z)=\arg a_0+\phi \Im\left(\frac{a_1}{a_0}\right)+\mathcal{O}(\phi^2),
\end{equation}

\noindent which yields:

\begin{align}
\Theta_0 &= \arg(a_0) - \psi_0=\theta,\\
\Theta_1 &= \Im\left( \frac{a_1}{a_0} \right)=-\frac{4\sin^2\theta}{2\cos\theta-3}\frac{M(\psi_0)}{D(\psi_0)},
\end{align}

\noindent where

\begin{align}\label{eq:CoefM}
    M(\psi_0)&=2-|w_0|\cos\psi_0-2\cos\theta \nonumber\\
    &+|w_0|\cos(\psi_0+\theta)+\cos(2\psi_0+\theta),
\end{align}

\noindent and

\begin{align}\label{eq:CoefD}
    D(\psi_0)&=-3+2\cos\theta-2|w_0|^2(1-\cos\theta)\nonumber\\
    &+2|w_0|(\cos\theta-1)\cos\psi_0-2|w_0|\sin\theta\sin\psi_0.
\end{align}

Assuming that the distance from the fixed point remains constant throughout the dynamics ($|w_n|=|w_0|=|w|$ for all $n$), the rotation number profile $\omega(|w|;\theta,\phi)$ can be obtained from averaging $\Theta/2\pi$ over the initial angle $\psi_0$. We have:

\begin{align}\label{eq:angular-average}
    \langle\Theta(|w|;\theta, \phi)\rangle&=\frac{1}{2\pi}\int_0^{2\pi}\left(\theta-\phi\frac{4\sin^2\theta}{2\cos\theta-3}\frac{M(\psi_0)}{D(\psi_0)}\right)\,d\psi_0,\nonumber\\
    \langle\Theta(|w|;\theta, \phi)\rangle&=\theta-\frac{\phi}{2\pi}\frac{4\sin^2\theta}{2\cos\theta-3}\int_0^{2\pi}\frac{M(\psi_0)}{D(\psi_0)}\,d\psi_0.
\end{align}

To solve the remaining integral, we define:

\begin{align}
    a&=-3+2\cos\theta-2|w|^2(1-\cos\theta),\\
    b&=2|w|(\cos\theta-1),\\
    c&=-2|w|\sin\theta,
\end{align}

\noindent then Eq. (\ref{eq:CoefD}) is rewritten as $D(\psi_0)=a+b\cos\psi_0+c\sin\psi_0$. Defining $d^2=b^2+c^2$ and using the right angle identities: $\sin\gamma=c/d$, $\cos\gamma=b/d$, and $\tan\gamma=c/b$, we perform a phase shift $\psi_0-\gamma=\varphi$ and obtain:

\begin{equation}\label{eq:CoefD-final}
    D(\varphi)=a+d\cos\varphi,
\end{equation}

\noindent which is an even function of $\varphi$. The same definitions are used to rewrite Eq. (\ref{eq:CoefM}). Discarding the terms proportional to $\sin(n\varphi)$, as they result in odd integrands and the corresponding integrals vanish, yields:

\begin{align}
    M(\varphi)&=2(1-\cos\theta)+|w|\cos(\theta-1)\cos\varphi\nonumber\\
    &+\cos\theta\cos2\varphi.
\end{align}

Thus, we have:

\begin{align}\label{eq:Integrals}
    \int_0^{2\pi}\frac{M(\psi_0)}{D(\psi_0)}\,d\psi_0&=\int_{0}^{2\pi}\frac{M(\varphi)}{D(\varphi)}\,d\varphi,\nonumber\\
    &=2(1-\cos\theta)\int_{0}^{2\pi}\frac{\,d\varphi}{a+d\cos\varphi}\nonumber\\
    &+2(\cos\theta-1)\int_{0}^{2\pi}\frac{\cos\varphi\,d\varphi}{a+d\cos\varphi}\nonumber\\
    &+\cos\theta\int_{0}^{2\pi}\frac{\cos2\varphi \,d\varphi}{a+d\cos\varphi}.
\end{align}

The integrals in Eq. (\ref{eq:Integrals}) are known:

\begin{equation}
    I_0=\int_{0}^{2\pi}\frac{\,d\varphi}{a+d\cos\varphi}=\frac{2\pi\sgn a}{\sqrt{a^2-d^2}},
\end{equation}
\begin{equation}
    I_n=\int_{0}^{2\pi}\frac{\cos{(n\varphi)}\,d\varphi}{a+d\cos\varphi}=I_0\left(\frac{-d}{a+\sqrt{a^2-d^2}\sgn{a}}\right)^n,
\end{equation}

\noindent and considering $|w|\ne0$, we have  $\sgn{a}=-1$. Thus, the final expression for the rotation number profile is

\begin{align}\label{eq:rotNumAnl}
    \omega(|w|;\theta,\phi) &= \frac{\theta}{2\pi}+\frac{\phi}{2\pi}\Omega(|w|;\theta),
\end{align}

\noindent where:

\begin{equation}\label{eq:OmegaAnl}
    \Omega(|w|;\theta)=\frac{2\sin^2(\theta/2)}{\Delta(\cos\theta-3)}\left(2K-\frac{d^2}{2(a-\Delta)}-\frac{d^2}{(a-\Delta)^2}\right),
\end{equation}

\noindent and:

\begin{align}
    K&=1-\cos\theta,\\
    d^2&=8K|w|^2,\\
    \Delta&=\sqrt{a^2-d^2}.
\end{align}

Figure~\ref{fig:fig2}(a) shows that, for sufficiently small $\phi$, the rotation number profile develops an extremum when $\theta\in(0,\pi/2)\cup(3\pi/2,2\pi)$, whereas it remains monotonic for all other angles.  This behavior is dictated by the $\theta$-dependence contained in Eq.(~\eqref{eq:OmegaAnl}).  To make this explicit we analyze Eq.(~\eqref{eq:OmegaAnl}) in the two asymptotic regimes $|w|\ll1$ and $|w|\gg1$. In each limit we expand the coefficients $a$ and $\Delta$ (see Table~\ref{tab:coeff}), substitute the results into Eq.~\eqref{eq:OmegaAnl}, and obtain

\begin{equation}\label{eq:leadingOmega}
    \Omega(|w|;\theta)= 
    \begin{cases}
\Omega_{0}(\theta)+\Omega_{2}(\theta)\,|w|^{2}+\mathcal{O}(|w|^{4}), & |w|\ll 1,\\[6pt]
-\dfrac{(2K+1)\sin^{2}\!\frac{\theta}{2}}{K\,(3-\cos\theta)}\,\dfrac{1}{|w|^{2}}
+\mathcal{O}\!\left(|w|^{-4}\right), & |w|\gg 1,
    \end{cases}
\end{equation}

\noindent with \(K=1-\cos\theta\) and coefficients  

\begin{equation}
    \Omega_0(\theta) = \frac{4 \sin^2\left(\frac{\theta}{2}\right) K}{(3 - \cos \theta)(3 - 2 \cos \theta)},
\end{equation}

\begin{equation}\label{eq:Omega2}
    \Omega_2(\theta)=\frac{-16\sin^2(\theta/2)(\cos\theta-1)^2}{(\cos\theta-3)(2\cos{\theta}-3)^3}\cos\theta.
\end{equation}

The term $\Omega_0(\theta)$ vanishes upon taking the $|w|$ derivative, thus not influencing its sign, and $\sgn(\Omega_2)=-\sgn(\cos\theta)$. Next, we evaluate the sign of the derivative $\partial\Omega/\partial |w|$ according to the leading–order expressions given by Eq.~(\ref{eq:leadingOmega}). For $|w|\gg1$ we find $\sgn{(\partial\Omega/\partial |w|)}=+1$ for all $\theta$, while for $|w|\ll1$ we find $\sgn{(\partial\Omega/\partial w)}=-\sgn(\cos\theta)$. Thus, $\partial\Omega/\partial |w|$ changes sign only if $\cos\theta>0$, justifying the $\theta$ interval for which an extremum $\Theta$ was observed numerically in the rotation number profile.

\begin{table}[h]
\caption{\label{tab:coeff}Leading–order coefficients entering Eq.~(\ref{eq:OmegaAnl}) in the
small- and large-\(|w|\) regimes, and $\sgn{(\partial\Omega/\partial |w|)}$.}
\begin{ruledtabular}
\begin{tabular}{ccc}
Quantity & $|w|\ll 1$ & $|w|\gg 1$ \\
\colrule
$a$          & $-3+2\cos\theta$ &
$\!\!-2K\,|w|^{2}$ \\
% $d^{2}$       & $8K\,w^{2}$ &
% $8K\,w^{2}$ \\
$\Delta$      & $3-2\cos\theta+\dfrac{4K}{-3+2\cos\theta}|w|^{2}$ &
$2K\,|w|^{2}$ \\
% $a-\Delta$    & $2(-3+2\cos\theta)-\dfrac{4K}{-3+2\cos\theta}w^{2}$ &
% $-\,4K\,w^{2}$ \\
% $\Omega(w;\theta)$ & $\Omega_{0}+\Omega_{2}w^{2}$ &
% $-\dfrac{(2K+1)\sin^{2}\!\frac{\theta}{2}}{K(3-\cos\theta)}\dfrac{1}{w^{2}}$ \\
$\sgn{(\partial\Omega/\partial |w|)}$&
$-\sgn(\cos\theta)$ &
$+1$ \\
\end{tabular}
\end{ruledtabular}
\end{table}

As our analytical expression for the frequency profile depends only on the action variable, since $|w|=\sqrt{2I}$, a Hamiltonian function $H=H_0+\phi H_1$ can be obtained by integrating Eq. (\ref{eq:rotNumAnl}) with respect to the action $I$. This procedure results in:
\begin{equation}\label{eq:AvgHam0}
    H_0(I;\theta)=-I\frac{\theta}{2\pi},
\end{equation}
which agrees with the Hamiltonian obtained for the integrable map in Appendix \ref{ap:action-angle} by choosing $\omega_0=\theta/2\pi$, and

\begin{align}\label{eq:AvgHam1}
    H_1(I;\theta)&=-\frac{1}{2\pi}\int_0^I\Omega_I(I;\theta)dI',\\
    &=\frac{3-2\cos\theta}{8\pi(3-\cos\theta)}\log\left(1+\frac{4K}{a+\Delta+2}\right),
\end{align}

\noindent where $\Omega_I(I;\theta)=\Omega(|w|=\sqrt{2I};\theta)$ is obtained from Eq. (\ref{eq:OmegaAnl}). $H_1$ is interpreted as an integrable perturbative Hamiltonian. The non-integrable angular dependency was averaged out by the integrals in Eq. (\ref{eq:angular-average}), to obtain the analytical approximation of the frequency profile.

\

\section{Numerical search for the fixed points}\label{ap:FixedPoints}

The integrable map in Eq. (\ref{eq:ikedaConsMatrixInt}) has only one fixed point, given by Eq. (\ref{eq:integrableFP}). For the nonintegrable map in Eq. (\ref{eq:ikedaCons}), the symmetry line $x=1/2$ can be used to find fixed points with $x^*=1/2$. The fixed point condition for Eq. (\ref{eq:ikedaCons}) becomes:

\begin{equation}\label{eq:FPcondition}
    f(y^*;\theta,\phi)=y^*-\frac{\sin{t_n}}{2(1-\cos{t_n})}=0, 
\end{equation}

\noindent where $t_n=\theta-\frac{\phi}{{y^*}^2+5/4}$. Thus, the fixed point calculation reduces to an one-dimensional root-finding problem. Multiple solutions are possible depending on the parameter values. For the parameter range considered in this work, it was sufficient to search for solutions in the interval $y\in{[-5,5]}$. This was done by dividing the interval into a grid of $1000$ points and looking for sign changes in $f(y;\theta,\phi)$. For each interval where a sign change was detected, the bisection method was used to refine the fixed point coordinate.

The stability of the solutions was verified by computing the corresponding residue \cite{Greene1978}, defined as $R=\frac{1}{4}(2-\text{tr} J)$, where the Jacobian matrix is evaluated at the fixed point. If $0<R<1$, the fixed point is stable; otherwise, it is unstable.

\

\begin{acknowledgments}
This study was supported by the Araucária Foundation, Brazil; the Coordination for the Improvement of Higher Education Personnel (CAPES), Brazil; the National Council for Scientific and Technological Development (CNPq), Brazil, under Grants Nos. 309670/2023-3 and 304616/2021-4; and the São Paulo Research Foundation (FAPESP), Brazil, under Grants Nos. 2019/07329-4, 2024/04557-4, 2024/14825-6 and 2024/05700-5.
\end{acknowledgments}

\

\section*{Data Availability Statement}

The data that support the findings of this study are openly available \cite{rep2025}. 

\

\section*{Author Declarations}

The authors have no conflicts to disclose.

% \bibliography{manuscript}% Produces the bibliography via BibTeX.


\end{document}